\documentclass[aps,10pt,pra,twocolumn,showpacs,letterpaper,showpacs,superscriptaddress]{revtex4-1}
\usepackage{graphicx,amsmath,amssymb,amsfonts,latexsym,color,dcolumn,bm,epsfig,subfigure,braket,mathdots,mathtools,tikz,array}

\usetikzlibrary{trees,arrows,shapes}\usetikzlibrary{decorations.pathmorphing}\usetikzlibrary{decorations.markings}\tikzset{partial ellipse/.style args={#1:#2:#3}{insert path={+ (#1:#3) arc (#1:#2:#3)}}}

\newcommand{\de}{\delta}\newcommand{\di}{\bm{d}}\newcommand{\e}{\varepsilon}\newcommand{\E}{\bm{E}}\newcommand{\f}{\bm{f}}\newcommand{\F}{\bm{F}}\newcommand{\G}{\tens{G}}\newcommand{\Gs}{\tens{G}^{(1)}}\newcommand{\Ga}{\Gamma}\newcommand{\hb}{\hbar}\newcommand{\hs}[1]{\hspace{#1}}\newcommand{\id}{\hs{-1ex}}\newcommand{\IM}{\T{Im}\,}\newcommand{\K}{\bm{k}}\newcommand{\kn}{_{kn}}\newcommand{\li}{\left}\newcommand{\m}{\mu}\newcommand{\mn}{_{mn}}\newcommand{\na}{\nabla}\newcommand{\nk}{_{nk}}\newcommand{\nm}{_{nm}}\newcommand{\Pa}{\bm{p}}\newcommand{\pa}{{||}}\newcommand{\q}{\quad}\newcommand{\qq}{\qquad}\newcommand{\R}{\bm{r}}\newcommand{\re}{\right}\newcommand{\RE}{\T{Re}\,}\newcommand{\s}{\bm{s}}\newcommand{\si}{\sigma}\newcommand{\T}[1]{\text{#1}}\newcommand{\TA}{\T{\fontsize{5}{5}\selectfont \,\id A}}\newcommand{\tens}[1]{\mbox{\textbf{\textsf{#1}}}}\newcommand{\V}{\bm{v}}\newcommand{\w}{\omega}

\renewcommand{\imath}[0]{\mathrm{i}}

\begin{document}

\title{Quantum Friction in Arbitrarily Directed Motion}

\author{J. Klatt} \affiliation{Physikalisches Institut, Albert-Ludwigs-Universit\"at Freiburg, Hermann-Herder-Str. 4, D-79104 Freiburg, Germany}
\author{M. Bel\'{e}n Far\'ias} \affiliation{Departamento de F\'isica, FCEyN, UBA and IFIBA, CONICET, Pabell\'on 1, Ciudad Universitaria, 1428 Buenos Aires, Argentina}
\author{D.A.R. Dalvit} \affiliation{Theoretical Divison, MS B213,  Los Alamos National Laboratory, Los Alamos, New Mexico 87545, USA}
\author{S.Y. Buhmann} \affiliation{Physikalisches Institut, Albert-Ludwigs-Universit\"at Freiburg, Hermann-Herder-Str. 4, D-79104 Freiburg, Germany}\affiliation{Freiburg Institute for Advanced Studies, Albert-Ludwigs-Universit\"{a}t Freiburg, Albertstr. 19, D-79104 Freiburg i. Br., Germany}
 
\date{\today}

\begin{abstract}
Quantum friction, the electromagnetic fluctuation-induced frictional force decelerating an atom which moves past a macroscopic dielectric body, has so far eluded experimental evidence despite more than three decades of theoretical studies. Inspired by the recent finding that dynamical corrections to such an atom's internal dynamics are enhanced by one order of magnitude for vertical motion -- compared to the paradigmatic setup of parallel motion -- we generalize quantum friction calculations to arbitrary angles between the atom's direction of motion and the surface in front of which it moves. Motivated by the disagreement between quantum friction calculations based on Markovian quantum master equations and time-dependent perturbation theory,  we carry out our derivations of the quantum frictional force for arbitrary angles employing both methods and compare them.
\end{abstract}

\maketitle

\begin{section}{Introduction}

The successful measurement of the Casimir-Polder force by, e.g., Sukenik \textit{et~al.} \cite{Sukenik1993} is one of the rather impressive investigations of the quantum vacuum and how it may be shaped through boundary conditions imposed by macroscopic bodies. As predicted by Casimir and Polder \cite{Casimir1948}, this force between a neutral atom and a polarizable macroscopic object arises from a position dependence in the atom's Lamb shift introduced by the presence of the body. A few decades after these early works theoreticians have come to agree that there must exist dynamical corrections to both the position-dependent Lamb shift and the resulting Casimir-Polder force when the atom moves parallel to the nearby surface \cite{Ferrell1980,Pendry1997,Volokitin2007}. Various theoretical methods have been employed, including time-dependent perturbation theory \cite{Barton2010,Barton2012}, quantum master equations \cite{Scheel2009}, generalized non-equilibrium fluctuation-dissipation relations \cite{Intravaia2014, Intravaia2016}, and influence-functional methods \cite{Farias2016}. Experimentally, however, such velocity-dependent corrections have eluded confirmation, as they are extremely small and short-ranged. In a recent work \cite{Klatt2016}, dynamical corrections to the internal dynamics, i.e., atomic level shifts and rates, were shown to be significantly larger when the atom moves vertically, rather than parallel to the macroscopic surface. And hence the question arises whether a similar qualitative change may be found for the quantum frictional force in the case of perpendicular motion -- potentially facilitating the experimental accessibility of the quantum friction force in such vertical setups. Hence, in this work, we generalize dynamical Casimir-Polder calculations from the paradigmatic scenario of parallel motion to arbitrarily directed motion.

Since the absence of an experimental benchmark has fostered the co-existence of alternative theoretical approaches to quantum friction -- some of them even giving contradictory predictions -- in this work we will carry out our derivations in a two-fold manner, using both Markovian quantum master equations and time-dependent perturbation theory. These two methods are known to lead to different results for the leading order in relative velocity of the quantum friction force for parallel motion, being given by a linear \cite{Scheel2009} and cubic \cite{Intravaia2015} dependency on velocity $\boldsymbol{v}$, respectively (we note that for certain exactly solvable models quantum friction for parallel motion has been shown to be cubic in velocity on asymptotic timescales \cite{Intravaia2016}). We find this discrepancy to prevail also for arbitrary angles between the atom's direction of motion and the surface.

For the parallel setup -- where it is possible to define a non-equilibrium steady state (NESS) in which the atom moves at constant velocity at a fixed distance from the plate -- generalized non-equilibrium fluctuation-dissipation relations can be employed in order to infer the dipole correlations entering the quantum friction force \cite{Intravaia2016}. By doing that, a recent study suggests that, in the asymptotic large time limit, the aforementioned discrepancies in the expressions for that force stem from different forms of the dipole power spectra implied using either approach \cite{Intravaia2016b}. The vanishing of the linear order in relative velocity seen in the fluctuation-relation approach is intimately linked to the symmetry of the atomic response function in the frequency domain. This symmetry -- more precisely, the Schwartz reflection principle required of any physical response function -- is \textit{necessarily} broken when the Markov approximation is applied, resulting in a non-vanishing linear order of the quantum friction force in atomic velocity.

However, for intermediate timescales -- that is, the realm of time-dependent perturbation theory and Markovian quantum master equations -- and moreover for the case of arbitrarily directed motion studied in this work, it is not possible to define such a NESS. In the former case, because stationarity is not yet reached, and in the latter because stationarity \textit{cannot} be reached at all since the atom is continuously approaching (or leaving) the surface. Hence, it is not possible to obtain an exact expression for the dipole spectrum, and therefore it is unclear which of two approaches for the quantum friction force acting upon an arbitrarily moving atom is appropriate, as they may simply apply in different temporal regimes.

We will study the paradigmatic setup of an atom moving next to a planar macroscopic body at zero temperature. The atom is neutral and possesses neither an electric nor a magnetic permanent dipole moment. The charge distribution provided by its constituents, however, is subject to quantum fluctuations which give rise to the atom's electric polarizability. The electric polarizability of the macroscopic body, which is assumed to be a metal or dielectric, is taken into account effectively through its permittivity $\varepsilon(\omega)$. The field induced by the atomic zero-point dipole fluctuations polarizes the body and thereby leads to the build-up of a mirror charge within it. That is, the field reflected by the surface is equal to a field induced by a fluctuating charge distribution, which for a perfect conductor is identical to the atom's but of opposite sign and mirrored at the interface. As the atom's dipole oscillates, the mirror atom's dipole does as well. Since both undulations are correlated, a finite dipole--dipole interaction emerges and hence an attractive force -- the Casimir-Polder force. The direction of the force is given by the axis between the atom and its mirror image. For an atom at rest in front of a plane surface, it points perpendicularly towards that surface and is nothing but the Casimir-Polder force $\F_\T{CP}$ \cite{Casimir1948}. As shown in Fig. \ref{fig:setup}, in case of a moving atom, fields induced by mirror images at previous times reach the actual atom and superpose in its current position, resulting in a tilted force. In addition to the ever-present Casimir-Polder force perpendicular to the surface, there then exists a finite force component pointing in the direction opposite to the atom's motion, causing it to decelerate. This force is what is called quantum friction in our context.

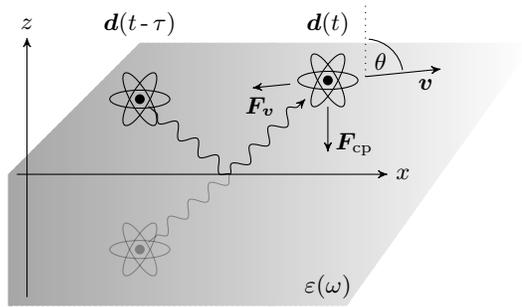
\begin{figure}
\begin{tikzpicture}[>=stealth',pos=.8,photon/.style={decorate,decoration=snake}]
\path[shade, left color = gray!70](-.25,-1.75)--(-.25,0)--(1.5,1.75) -- (6.75,1.75) -- (4.25,-1.75) -- cycle;
\node[black] at (4.,-1.5) {$\varepsilon(\w)$};
\node[] (xbegin) at (-.25,0)    {}; \node[] (xend) at (5,0) {$x$}; \draw[->] (xbegin)--(xend);
\node[] (zbegin) at (0,-1.75)   {}; \node[] (zend) at (0,2) {$z$}; \draw[->] (zbegin)--(zend);
\node[] (atom1)  at (1.5,1.)  {};
\draw[fill=black,draw=none] (1.5,1.) circle (.065);
\draw[] (1.5,1.) [partial ellipse=0:360:.4cm and .12cm];
\draw[rotate around ={120:(1.5,1.)}] (1.5,1.) [partial ellipse=0:360:.4cm and .12cm];
\draw[rotate around ={-120:(1.5,1.)}] (1.5,1.) [partial ellipse=0:360:.4cm and .12cm];
\node[] at (1.5,2.) {$\bm{d}(t\,\T{-}\,\tau)$};
\node[] (atom2)  at (4,1.25)  {};
\draw[fill=black,draw=none] (4,1.25) circle (.065);
\draw[] (4,1.25) [partial ellipse=0:360:.4cm and .12cm];
\draw[rotate around ={120:(4,1.25)}] (4,1.25) [partial ellipse=0:360:.4cm and .12cm];
\draw[rotate around ={-120:(4,1.25)}] (4,1.25) [partial ellipse=0:360:.4cm and .12cm];
\node[] at (4,2.) {$\bm{d}(t)$};
\node[opacity=.333] (mirror) at (1.5,-1.) {};
\draw[fill=black,draw=none,opacity=.333] (1.5,-1.) circle (.065);
\draw[opacity=.333] (1.5,-1.) [partial ellipse=0:360:.4cm and .12cm];
\draw[opacity=.333,rotate around ={120:(1.5,-1.)}] (1.5,-1.) [partial ellipse=0:360:.4cm and .12cm];
\draw[opacity=.333,rotate around ={-120:(1.5,-1.)}] (1.5,-1.) [partial ellipse=0:360:.4cm and .12cm];
\draw[-,photon,opacity=.333] (2.7,0)--(mirror);
\draw[->,photon,](2.7,0)--(3.7,1);
\draw[-,photon] (2.7,0)--(atom1);
\draw[dotted] (4.5,1.3)--(4.5,2.25);
\draw[] (4.5,1.3) [partial ellipse=10:85:.5cm and .5cm];\node[] at (4.7,1.5) {$\theta$};
\draw[->] (4.5,1.3)--(5.5,1.4);\node[] at (5.3,1.2) {$\bm{v}$};
\draw[->] (3.5,1.2)--(3,1.15);\node[] at (3.1,.9) {$\bm{F_v}$};
\draw[->] (4,.9)--(4,.3);\node[] at (4.35,.4) {$\bm{F}_\T{cp}$};
\end{tikzpicture}
\caption{Atom moving next to a surface with velocity $\V$ at zero temperature. Its electric dipole $\di(t)$ fluctuates about zero, resulting in emission of virtual photons. The atom may, e.g., have emitted a photon at time $t-\tau$ which after reflection is reabsorbed at time $t$. This interaction leads to the Casimir-Polder force $\F_\T{CP}$ attracting the atom towards the surface, as well as a frictional force $\F_{\V}$ counter-acting the relative motion.}
\label{fig:setup}
\end{figure}

The paper is organized as follows. In Section II the common ground for both the Markovian and the perturbative approaches to quantum friction is layed out. Subsequently, in Section III, we develop the Markovian quantum master equation approach to quantum friction in arbitrarily directed motion, while in Section IV we contrast this calculation with time-dependent perturbation theory. Lastly, Section V contains our conclusions.
\end{section}

\begin{section}{Setup}\label{sec:setup}

As mentioned in the Introduction, we here consider an atom moving in the proximity of a homogeneous, dielectric, half space, $z<0$, while the atom itself is placed in vacuum at zero temperature. The Hamiltonian of the entire system consists of atomic, field and interaction contributions,
\begin{align}\label{eq:hamiltonian}
\hat H=
&\hat H_\T{A}+\hat H_\T{F}+\hat H_\T{AF}\\\label{eq:atom}=
&\frac{\hat \Pa_\TA^2}{2m_\TA}+\sum_{n=0}^\infty E_n \hat A_{nn}\\\label{eq:field}
 &+\hb\sum_{\sigma=\T{e},\T{m}}\int\!\! d\R\id\int_0^\infty\!\!\id\id d\w\w\,\hat \f^\dag_\si(\R,\w)\,\,\id\cdot\,\id\hat \f_\si(\R,\w)\\\label{eq:int}
 &-\sum\mn \hat A\mn \di\mn\,\,\id\cdot\,\id\hat \E(\R_\TA).
\end{align}
Here, $\hat \Pa_\TA$ is the atom's center-of-mass momentum operator and the $E_n$ are the atom's internal eigenenergies. The $\hat A\mn=\ket{m}\bra{n}$ are so-called flip operators which, for $m\,\id=\,\id n$, project onto the $n^\T{th}$ eigenstate or, for $m\,\id\neq\,\id n$, induce transitions from state $\ket{n}$ to $\ket{m}$.

The medium-assisted excitations $\hat \f^\dag_\si$ result from the quantization of the field in the presence of the bulk medium \cite{Huttner1992,Philbin2010}:
\begin{align}
\hat \E(\R)=\id\sum_{\sigma=\T{e},\T{m}}\int\!\! d\R'\id\int_0^\infty\!\!\id\id d\w\,\tens{G}_\si(\R,\R',\w)\,\,\id\cdot\,\id\f_\si(\R',\w)+\T{h.c.}.
\end{align}
They can be thought of as representing electric (e) or magnetic (m) unit dipoles residing in $\R'$ and oscillating at frequency $\w$, thereby populating the appropriate field mode. They act upon the field's vacuum state $\ket{\{0\}}$ as
\begin{align}
\hat\f_\si(\R,\w)\ket{\{0\}}=&\,0,\\
\hat \f^\dag_\si(\R,\w)\ket{\{0\}}=&\ket{\bm{1}_\si(\R,\w)},
\end{align}
and fulfill bosonic commutation relations. The electric and magnetic coefficients,
\begin{align}
&\tens{G}_\T{e}(\R,\R',\w)=i\tfrac{\w^2}{c^2}\sqrt{\tfrac{\hb\e_0}{\pi}\T{Im}\,\e(\w)}\,\tens{G}(\R,\R',\w)\,,\\
&\tens{G}_\T{m}(\R,\R',\w)=i\tfrac{\w}{c}\sqrt{\tfrac{\hb}{\pi\m_0}\tfrac{\T{Im}\,\m(\w)}{|\m(\R',\w)|^2}}[\na'\times\tens{G}(\R,\R',\w)]^\T{T},
\end{align}
respectively, derive from the electromagnetic Green's tensor $\tens{G}$. The latter is defined as the formal solution to the homogeneous Helmholtz equation which arises from Faraday's and Amp\`ere's law, supplemented by vanishing boundary conditions at $|\R-\R'|\to\infty$ \cite{Buhmann2013}. The $\tens{G}_\si$ fulfill the integral relation
\begin{align}\label{eq:intrel}
\sum_\si\,\id\int\id d\s\tens{G}_\si(\R\!,\s,\w)\!\!\cdot\!\!\tens{G}_\si^*(\s,\R'\!\!\!,\w)=\tfrac{\hb\m_0\w^2}{\pi}\T{Im}\tens{G}(\R\!,\R'\!\!\!,\w).
\end{align}
Calculating the variance of $\hat\E$ using this relation and invoking the fluctuation-dissipation theorem reveals that $\mu_0\w^2\tens{G}$ is the linear response function of the electromagnetic field. In order to eventually evaluate the force, the Green's tensor has to be specified according to the geometry and material properties of the bulk medium. We will adhere to a Drude-Lorentz modeled dielectric and the non-retarded limit of the half-space scattering Green's tensor \cite{Buhmann2013},
\begin{align}\label{eq:Gs2}
\Gs\,\id(\R,\R'\,\id,\w)=\tfrac{r_\T{p}(\w)c^2}{8\pi^2\w^2}\!\!\int\!\!\tfrac{d^2\K^\pa}{k^\pa}\,(\K\otimes\K^*)\,e^{i\K^\pa\cdot(\R-\R')-k^\pa(z+z')}.
\end{align}
It describes the near-field scattering of medium-assisted field excitations of frequency $\w$ and wave vector
\begin{align}
\K=(\K^\pa,ik^\pa)=(k^\pa\cos\phi,k^\pa\sin\phi,ik^\pa)\,.
\end{align}
The reflection is governed by the Fresnel reflection coefficient $r_\T{p}(\w)$ for transverse magnetic radiation of frequency $\w$, whose non-retarded limit reads
\begin{align}
r_\T{p}(\w)=\frac{\e(\w)-1}{\e(\w)+1}\,.
\end{align}
The permittivity entering the reflection coefficient should be local and dispersive for our results to hold.

The last term of the Hamiltonian (\ref{eq:hamiltonian}) describes the interaction of atom and field in electric-dipole approximation where the atomic dipole $\di$ has been expanded in the flip-operator basis. The force acting upon the atom is
\begin{align}\label{eq:force}
\F(t)=\langle\hat\di\cdot\nabla\hat\E(\R_\TA)\rangle\,.
\end{align}
As was already studied in Ref.~\cite{Intravaia2015}, the force crucially depends on the atom's trajectory. In the Markovian approach (Section III), the trajectory can be assumed approximately straight and uniform
\begin{align}
\R_\TA -\R'_\TA \simeq\V_\TA(t-t'),
\end{align}
on time scales of the field's auto-correlation time. Within the perturbative approach (Section IV), we will assume that the atom is at rest for times $t<0$ and that it moves at constant velocity $\V_\TA=v(\sin\theta,0,\cos\theta)$ with $v>0$ for times $t>0$. This sudden boost trajectory is precisely that used in Ref.~\cite{Barton2010} in the case of parallel motion. Note that this prescribed uniform motion for $t>0$ is maintained by an external force counteracting the quantum friction force. For $t<0$ the atom is located at $\R_\TA(t<0)=(x_0,y_0,z_0)$ and for $t>0$ its trajectory is given by
\begin{equation}
\R_\TA(t)=(x_0+vt\sin\theta,y_0,z_0+v t \cos\theta).
\end{equation}
Note that $\theta\!=\!\pm\pi/2$ corresponds to parallel motion, $\theta\!=\!\pi$ to vertical motion towards the plane, and $\theta\!=\!0$ indicates vertical motion away from the plane. Trajectories containing, e.g., a continuous acceleration from zero velocity to constant final velocity over a given time interval, have been considered in Ref.~\cite{Intravaia2015} for the parallel motion case, and could be analogously implemented for our case of arbitrarily directed motion. However, for simplicity, in this paper we only consider the sudden boost trajectory described above.
\end{section}

\begin{section}{Markovian Approach}\label{sec:markov}

We will first solve the internal dynamics of the atom by means of a Markovian quantum master equation. This then further allows for inferring atomic dipole correlations via the quantum regression hypothesis \cite{Lax2000}, which can be proven to hold as a theorem for semi-group Markovian processes \cite{Breuer2002}. The such obtained dipole correlations eventually lead to an evaluable expression of the quantum friction force.

\begin{subsection}{Internal Atomic Dynamics}

By means of a Born-Oppenheimer type argument, the internal dynamics of the atom can be separated from its center-of-mass motion. That is, it can be solved for an arbitrary but fixed instantaneous position $\R_\TA$ and momentum $\Pa_\TA$. Subsequently, the force determining the change in this very momentum may be calculated for given internal dynamics. The latter are captured by the time evolution of the flip operators $\hat A_{mn}$. In this subsection we solve the corresponding Heisenberg equation, showing that it leads to velocity-dependent rates of spontaneous decay and eigenfrequencies, respectively. Note that the full Hamiltonian includes the field surrounding the atom. This environment will be traced over later on, leading to dissipative dynamics in the first place.

At any point in time, the full Hamiltonian can be decomposed as in (\ref{eq:hamiltonian}). The purely atomic operators $\hat A\mn$ then commute with the field contribution (\ref{eq:field}) at equal times, since they live in orthogonal subspaces of the total Hilbert space. The commutator with the atomic Hamiltonian (\ref{eq:atom}) yields the eigenfrequencies as in the absence of the field, while the commutator with the interaction Hamiltonian (\ref{eq:int}) will lead to the Lamb shift and Einstein rates:
\begin{align}\label{eq:HeisA}
 \dot{\hat A}\mn(t)=i\w\mn \hat A\mn+\tfrac{1}{i\hb}[\hat A\mn(t),\hat H_\T{AF}(t)].
\end{align}
This differential equation may be solved formally and re-substituted into itself \textit{ad infinitum}. This procedure results in a Dyson-like expansion based on which one may design a series of approximate solutions $\hat A\mn^{(k)}$, where $k$ is even, converging towards the exact one as $k\,\id\rightarrow\,\id\infty$. The $k^\T{th}$ element of the above series is of order $\di^{k}$. Therefore, its dynamics comprise re-absorption of up to $k/2$-fold reflected photons. In the following, multiple reflections by the atom will be neglected. That is, the dynamics will be solved up to an order $k$ equal to two. Employing the prescription above, one writes
\begin{align}\label{eq:a1}
 \dot{\hat A}\mn^{(2)}(t)=&\,\dot{\hat A}\mn^{(0)}(t)+\tfrac{1}{i\hb}[\hat A\mn^{(0)}(t),\hat H_\T{AF}^{(2)}(t)]\q,\\\label{eq:h0}
 \hat H_\T{AF}^{(2)}(t)=&\,-\sum\mn \hat A\mn^{(0)}(t)\di\mn\,\,\id\cdot\,\id\hat \E^{(1)}(\R_\TA,t).
\end{align}
Beware that $\hat \E^{(1)}$ denotes the free field plus the field induced by an atom described as $\hat A\mn^{(0)}$. Starting from the Heisenberg equation for the $\hat \f_\si$ and employing the integral relation (\ref{eq:intrel}), one arrives at
\begin{align}\label{eq:e0}
 \hat \E^{(1)}(\R_\TA,t)=
 &\sum_\si\int\!\! d\R\id\int_0^\infty\!\!\id\id d\w\,\tens{G}_\si(\R_\TA,\R,\w)\cdot\hat \f_\si(\R,\w)\\\nonumber
 &+\tfrac{i\m_0}{\pi}\sum\mn\int_{t_0}^t\id\!dt'\id\int_0^\infty\!\!\id\id d\w\w^2e^{i\w\mn(t-t')}\\\nonumber
 &\times \T{Im}\tens{G}(\R_\TA,\R'_\TA,\w)\,\,\id\cdot\,\,\id\di\mn \hat A^{(0)}\mn(t')+\T{h.c.} ,
\end{align}
where absence of a time argument indicates initial time $t_0$, and primed quantities are evaluated at $t'$. Substituting the above field into (\ref{eq:h0}) makes it apparent that the interaction process taken into account is re-absorption of a reflected, medium assisted,  photon that has been emitted by the atom at time $t'$ in history, i.e., a dipole interaction between $\di(t)$ and $\di(t')$. Each of these reflected photons accumulates a phase $i\w\mn(t-t')$ as it travels from the place of its creation $\R'_\TA$ to its destination $\R_\TA$, where it eventually interferes with others of its kind, according to their respective relative phases. Inserting (\ref{eq:e0}) into the normal-ordered vacuum-expectation value of (\ref{eq:a1}) leads to the reduced dynamics of the atom's internal:
\begin{align}\nonumber
&\langle\dot{\hat A}^{(2)}\mn(t)\rangle=\,i\w\mn\langle\hat A^{(2)}\mn(t)\rangle\\\label{eq:master}
&\qq\qq\qq-\sum_k[C_{nk}(t)+C^*_{mk}(t)]\langle\hat A^{(2)}\mn(t)\rangle,
\end{align}
with the coefficients
\begin{align}\nonumber
&C\nk=\,\tfrac{\m_0}{\pi\hb}\int_{t_0}^t\id\!dt'\id \int_0^\infty\!\!\id\id d\w\w^2\,e^{-i(\w-\w\nk)(t-t')}\\\label{eq:c}
&\qq\qq\qq \times \di\nk\cdot\T{Im}\tens{G}(\R_\TA,\R'_\TA,\w)\cdot\di\kn.
\end{align}
We here employed the fact that in the absence of degenerate and quasi-degenerate dipole-transitions in the atom, the off-diagonal flip-operator dynamics effectively decouple from the diagonal ones as well as from each other due to the orthogonality of non-commensurate oscillations and neglected terms of order $d^4$.

Now, the real part of the $C\nk$ renders the rates $\Ga_n$ of spontaneous decay of $\ket{n}$, while their imaginary part delivers the corrections $\hbar\de\w_n$ to the eigenenergy $E_n$ of the free atom -- both stemming from the interaction with the field:
\begin{align}\label{eq:dw}
&\de\w_n=\sum_k\de\w\nk=\sum_k\IM C\nk\,,\\\label{eq:G}
&\Ga_n=\sum_k\Ga\nk=2\sum_k\RE C\nk\,.
\end{align}
The $t'$ integration in (\ref{eq:c}) involves the Green's tensor via $\R'_\TA$, and the oscillating exponential describing a field excitation of frequency $\w\nm$ traveling from $\R'_\TA$ to $\R_\TA$. In the spirit of the Doppler effect, the spatial distance between the photon's origin and destination may equally well be translated into a shift in frequency. Under the conditions of (a) uniform motion on timescales of the field's auto-correlation time and (b) $t$ much larger than these times, performing the limit of $t_0\to-\infty$, i.e., a Markov approximation, and employing the non-retarded Green's tensor (\ref{eq:Gs2}) leads to
\begin{align}\label{eq:cres}
&C\nk^\T{res}\!=\!-\tfrac{i}{8\pi^2\hb\e_0}\id\int_0^{2\pi}\id\id\id d\phi\!\!\int_0^\infty\id\id\!\!dk^\pa k^{\pa2} d\nk^{( \phi)2}r_\T{p}(\w'\nk)\Theta(\w'\nk)e^{-2k^\pa z_\TA(t)}\!\!,\\\label{eq:cnres}
&C\nk^\T{nres}\!=\!\tfrac{i}{8\pi^3\hb\e_0}\id\int_0^\infty\id\id\!\!d\xi\!\!\int_0^{2\pi}\id\id\id d\phi\!\!\int_0^\infty\id\id\!\!dk^\pa k^{\pa2} d\nk^{( \phi)2}\frac{\w'\nk r_\T{p}(i\xi)}{(\w\nk^{\prime2}+\xi^2)}e^{-2k^\pa z_\TA(t)},
\end{align}
for the resonant and non-resonant contributions to the Heisenberg coefficients (\ref{eq:c}), respectively \cite{Klatt2016}. Above we have introduced the shorthand notation,
\begin{align}
& d\nk^{(\phi)2}=\di\nk\cdot\li(\id\begin{smallmatrix}\cos^2\phi&\cos\phi\sin\phi&-i\cos\phi\\\cos\phi\sin\phi&\sin^2\phi&-i\sin\phi\\i\cos\phi&i\sin\phi&1\end{smallmatrix}\,\id\re)\cdot\di\nk\,,
\end{align}
and the Doppler-shifted, complex-valued, frequency
\begin{align}\label{eq:wtilde}
\w'\nk=\w\nk+vk^\pa(\sin\theta\cos\phi-i\cos\theta)\,.
\end{align}
The Heaviside step function appearing in Eq.~(\ref{eq:cres}) is understood with respect to the real part of $\w'\nk$. Without loss of generality, the coordinate system was chosen such that the $y$-component of the velocity vanishes. The direction of the atomic velocity $\V$ is hence solely determined by the angle $\theta$ between $\V$ and the $z$-axis. Depending on $\theta$, the frequency $\w\nk$ may be shifted along the imaginary axis, which is unusual. The Doppler effect can be understood as the shortened/lengthened time interval between the passing of two consecutive wave fronts through a certain (and possibly itself moving) observation point in space due to the relative motion of the waves' source and that very point. In the scenario at hand, the source is the atomic dipole at a time $t$-$\tau$ in the past and the observation point is the position of the atom at time $t$. While in the case of the atom moving parallel to the surface, there are actual wave fronts propagating along the direction of motion, in the case of vertical motion, in the non-retarded regime, only evanescent waves perpetuate along the direction of motion. The latter do not possess such thing as a wave front and thus the traditional intuition for the Doppler effect does not apply. Just as evanescent waves are characterized by a complex $k$-vector, the Doppler shift we encounter in vertical motion has a component along the imaginary axis. Finally, we emphasize that since we are using the non-retarded form of the Green's tensor, all the results below will be valid as long as the atom is always within the near-field zone.

The expressions (\ref{eq:cres}) and (\ref{eq:cnres}) for the resonant and non-resonant contributions to the internal dynamics of an atom moving in front of a half-space hold for arbitrarily directed atomic motion. However, since they result from a series expansion of convergence radius $v/z_\TA(t)\,\w\nk$ (see Ref.~\cite{Klatt2016}) a condition which has to be fulfilled is $v\!<\!z_\TA(t)\w\nk$. The static Casimir-Polder shifts and rates are re-obtained by replacing the primed, i.e., Doppler-shifted, frequencies $\w'\nk$ in (\ref{eq:cres}) and (\ref{eq:cnres}) by their bare counterparts $\w\nk$, which is identical to considering only the zeroth order in relative velocity. For finite velocity, the integrals in (\ref{eq:cres}) and (\ref{eq:cnres}) can be solved numerically. In the case of parallel motion, due to the rotational symmetry of the setup, shifts and rates only vary quadratically with the atom's velocity. As for vertical motion, the variation is linear \cite{Klatt2016}. Therefore, for small velocities the dynamical corrections to the static shifts and rates are significantly larger if the atom moves perpendicularly to the surface.

Since they are needed later on, we will conclude this Section by giving both the decay rate and non-resonant level shift for a ground-state atom as prescribed by the real and imaginary part of  \eqref{eq:cnres}. In order to match the configuration of the perturbative calculation, the atom is assumed to be a two-level system. Hence, for an isotropic ground state atom,
\begin{align}\nonumber
\Ga_0^{(\theta=\frac{\pi}{2})}
&=\tfrac{d^2}{2\pi^2\hb\e_0}\int_0^\infty\id\id\!\!d\w\!\!\int_0^\infty\id\id\!\!d^2k^\pa k^\pa e^{-2k^\pa z_0}\IM r_p(\w)\\\label{eq:Ga01res}
&  \qq \times \de(\w+\w_{10}-k^\pa v\cos\phi), \,\\\nonumber
\Ga_0^{(\theta\neq\frac{\pi}{2})}
&\simeq-\tfrac{3 d^2v\cos\theta}{8\pi^2\hb\e_0z_\TA^4(t)}\int_0^\infty\id\id\!\!d\w\;\tfrac{\IM r_p(\w)}{(\w+\w_{10})^2}\\\label{eq:Ga01nres}
&=\tfrac{3d^2v\cos\theta}{8\pi^2\hb\e_0z_\TA^4(t)}\int_0^\infty\id\id\!\!d\xi\;\tfrac{\xi^2-\w_{10}^2}{(\xi^2+\w_{10}^2)^2}\;r_p(i\xi), \,\\\nonumber
\de\w_0
&\simeq-\tfrac{d^2}{8\pi^2\hb\e_0z_\TA^3(t)}\int_0^\infty\id\id\!\!d\w\;\tfrac{\IM r_p(\w)}{\w+\w_{10}}\left[1-\tfrac{3}{4}\tfrac{v^2(1+3\cos2\theta)}{(\w+\w_{10})^2z_\TA^2(t)}\right]\\\label{eq:dw01}
&=-\tfrac{d^2}{8\pi^2\hb\e_0z_\TA^3(t)}\int_0^\infty\id\id\!\!d\xi\;\tfrac{\w_{10} }{\w_{10}^2+\xi^2} r_p(i\xi) \\\nonumber
&\qq\qq\qq \times \left[1-\tfrac{3}{4}\tfrac{v^2(\w_{10}^2-\xi^2)(1+3\cos2\theta)}{(\w_{10}^2+\xi^2)^2z_\TA^2(t)}\right] .
\end{align}
Note that the rates signify a spontaneous excitation of the atom. In the parallel case, however, this process is resonant. Due to the resonance-enforcing Heaviside-$\Theta$ function, the probability for this Cherenkov-type excitation is only finite if the Doppler-shift energy $\hb vk^\pa$ is large enough to bridge the gap between the ground state and first excited state. This imposes a constraint on the parallel projection of the wave vector, namely $k^\pa>\w_{10}/v$. This constraint manifests itself in the exponential of the $k$-space integration in both the rate and the shift and results in an exponential suppression of the motion-induced excitation of a parallely moving ground-state atom as the speed of the latter approaches zero \cite{Intravaia2015}. Not so in non-parallel motion, where to leading order the excitation rate scales linearly with velocity. Remarkably, as for the non-resonant level shift, the directionality does not change the leading-order power-law in $v$. It scales quadratically with atomic velocity in any case. Lastly, note that for parallel motion the above expressions perfectly coincide with the ones obtained perturbatively in \cite{Barton2010,Intravaia2015} if one recalls that it needs a factor $4\pi\e_0$ in order to translate from SI units (used in our work) to Gaussian units (used in \cite{Barton2010,Intravaia2015}).
\end{subsection}

\begin{subsection}{Casimir-Polder and Friction Force}\label{sec:comm}

After having solved the internal atomic dynamics for an arbitrary but fixed atomic center-of-mass velocity, one can now solve the Newtonian dynamics of the atom for arbitrary but fixed atomic transition frequencies and decay rates. The force (\ref{eq:force}) determining the Newtonian dynamics of the atom can be decomposed into its projection $F_v$ onto the atom's direction of motion and a component orthogonal to that. While the former decelerates the atom, the latter changes its direction of motion. Since we are interested in a frictional force, we study the gradient $\nabla_v\equiv \V\cdot\nabla/v$ of the force along the atom's velocity. The electric field entering the force (\ref{eq:force}) was already determined in (\ref{eq:e0}). Inserting it and normally-ordering atom and field operators leads to
\begin{align}\label{eq:F1}
F_v(t)&=\frac{i\mu_0}{\pi}\int_0^\infty\id\id\!\!d\tau\id\int_0^\infty\id\id\!\!d\w\w^2e^{-i\w\tau}\\\nonumber
&\times \nabla_v\langle\di(t)\cdot\IM\G^{(1)}(\R_\TA,\R'_\TA,\w)\cdot\di(t-\tau)\rangle+\T{h.c.}\,.
\end{align}
This clearly shows the dependence of the force on the auto-correlation of the atomic dipole. Within the Markov approximation, one may infer the two-point correlation function needed above via Lax's quantum-regression hypothesis \cite{Lax2000,Breuer2002}. It gives
\begin{align}\label{eq:C}
\langle\di(t)\di(t\!-\!\tau)\rangle\!=\!\sum_{nk}\!\di_{nk}\di_{kn}\,p_n(t)\,e^{[i\w\nk-\frac{1}{2}(\Ga_n+\Ga_k)]\tau},
\end{align}
where $p_n(t)\!\equiv\!\langle A_{nn}(t)\rangle$ is the population of the atomic state $\ket{n}$. Inserting Eq.~(\ref{eq:C}) into the force (\ref{eq:F1}), and restricting to the non-retarded regime, leads to a decomposition of the friction force into contributions associated with the various energy levels of the atom, each weighted with the population $p_n$ of that level,
\begin{align}
F_v(t)=\sum\nk p_n(t)F\nk(t)\,.
\end{align}
The summands are given by
\begin{align}\label{eq:F}
&F\nk(t)=-\tfrac{1}{4\pi^3\e_0}\int_0^\infty\id\id\!\!d\w\id\int_0^{2\pi}\id\id\!\!d\phi\id\int_0^\infty\id\id\!\!dk^\pa k^{\pa3}e^{-2k^\pa z_\TA(t)}\, d\nk^{(\phi)2}\\\nonumber
&\times \frac{(\w+\Omega'\kn)\cos\theta+\frac{1}{2}(\Ga'_n+\Ga'_k)\sin\theta\cos\phi}{(\w+\Omega'\kn)^2+\frac{1}{4}(\Ga'_n+\Ga'_k)^2}\,\T{Im}r_{\,\,\id\T{p}}(\w)\,,
\end{align}
 where the capital omega indicates that frequencies include the formerly calculated Casimir-Polder shifts (\ref{eq:dw}),
\begin{align}
\Omega\nk=\w\nk+\de\w_n-\de\w_k\,,
\end{align}
and the primed shifts and rates carry Doppler shifts 
\begin{align}\label{eq:Omprime}
\Omega\nk'
&=\Omega\nk-vk^\pa\sin\theta\cos\phi\,,\\\label{eq:Gaprime}
\Ga\nk'
&=\Ga\nk-vk^\pa\cos\theta\,.
\end{align}

We will now focus on an atom which at time $t$ is in its ground state, i.e., 
$p_n(t)=\delta_{n,0}$, and only consider its first excited state when calculating the Casimir-Polder and quantum friction force, i.e., $F_v(t)=F_{01}(t)$. This facilitates the comparison to the perturbative calculations of Section~IV. For better readability, we will drop subscripts, that is $\Omega\!\equiv\!\Omega_{10}$ and $\Ga\!\equiv\!\tfrac{1}{2}(\Ga_0+\Ga_1)$. The full force acting on the ground-state atom can be split into a resonant term -- which stems from the pole in the $\w$-integration -- and a non-resonant term. This decomposition gives
\begin{align}\nonumber
&F_v^\T{res}(t)
=-\tfrac{1}{4\pi^2\e_0}\,\T{Re}\int_0^{2\pi}\id\id\!\!d\phi\id\int_0^\infty\id\id\!\!dk^\pa k^{\pa3}e^{-2k^\pa z_\TA(t)}\,d^{(\phi)2}\\\label{eq:Fres}
&\times (\cos\theta-i\sin\theta\cos\phi)\,r_{\,\,\id\T{p}}(-\Omega'+i\Ga')\,\Theta(-\Omega'),\\\nonumber
&F_v^\T{nres}(t)
=-\tfrac{1}{4\pi^3\e_0}\,\T{Re}\!\!\int_0^{2\pi}\id\id\!\!d\phi\id\int_0^\infty\id\id\!\!dk^\pa k^{\pa3}e^{-2k^\pa z_\TA(t)}\,d^{(\phi)2}\\\label{eq:Fnres}
&\times (\cos\theta-i\sin\theta\cos\phi)\id\int_0^\infty\id\id\!\!d\xi\,\frac{\Omega'-i\Ga'}{(\Omega'-i\Ga')^2+\xi^2}\,r_{\,\,\id\T{p}}(i\xi)\,,
\end{align}
respectively. Again, due to the Heaviside-$\Theta$ function, the resonant force is only finite if the Doppler-shift energy $\hb vk^\pa$ is large enough to bridge the gap between the ground state and  first excited state. The resonant friction force hence stems from a Cherenkov-like excitation of the atom and is exponentially suppressed due to the constraint on $k^\pa$ imposed by the $\Theta$ function \cite{Intravaia2015}. We will therefore neglect that term and focus on the non-resonant friction force in the following. The velocity dependence of that non-resonant force (\ref{eq:Fnres}) is of two-fold nature. The $v$-dependence brought about by the Doppler-shifted frequencies $\Omega'$ and rates $\Ga'$  we will call \textit{explicit} dependence. The velocity dependence $\de^vC\nk$ already included in the non-shifted $\Omega$ and $\Ga$ via the coefficients
\begin{align}
C\nk=\left.C\nk\right|_{v=0}+\de^vC\nk\,,
\end{align}
given in (\ref{eq:c}), instead, will be referred to as \textit{implicit} dependence. Eventually, up to linear order in $v$ the force acting upon the atom reads,
\begin{align}\label{eq:FMarkov0}
&F_v(t)\simeq-\frac{3d^2}{8\pi^2\e_0z_\TA^4(t)}\int_0^\infty\id\id\!\!d\w\,\frac{(\w+\Omega)\,\IM r_{\,\,\id\T{p}}(\w)}{(\w+\Omega)^2+\Ga^2}\\\nonumber
&\times\!\li[\!\cos\theta\!+\!\frac{2v\,\Ga(1+\cos^2\!\theta)}{z_\TA(t)[(\w+\Omega)^2+\Ga^2]}\!-\!\li(\de^vC^*_{01}+\de^vC_{10}\re)\cos\theta\re]\\\nonumber
&\qq=-\frac{3d^2}{8\pi^2\e_0z_\TA^4(t)}\int_0^\infty\id\id\!\!d\xi\,\frac{\Omega(\Omega^2+\Ga^2+\xi^2)\,r_{\,\,\id\T{p}}(i\xi)}{(\Omega^2+\Ga^2-\xi^2)^2+4\Omega^2\xi^2}\\\nonumber
&\times\!\li[\!\cos\theta\!+\!\frac{2v\,\Ga(1+\cos^2\!\theta)}{z_\TA(t)[(\w+\Omega)^2+\Ga^2]}\!-\!\li(\de^vC^*_{01}+\de^vC_{10}\re)\cos\theta\re]\!.
\end{align}
The first summand gives the static Casimir-Polder force's projection onto the direction of motion. The second summand comprises the explicit velocity dependence and the last one stems from the implicit dependence mentioned above.

As an anchor to previous results as well as the perturbative calculations to come, we now start by deducing the static Casimir-Polder force up to order $d^2$ from the expression \eqref{eq:FMarkov0}. This is done by omitting the decay rate as well as the corrections $\de\w_0$ to the bare frequency $\w_{10}$, and setting the velocity to zero. For an isotropic preparation of the atom, i.e., $d_x\!=\!d_y\!=\!d_z\!\equiv\!d$, this yields 
\begin{align}\nonumber
F^{(2)}_\T{CP}(t)
&=-\frac{3d^2\cos\theta}{8\pi^2\e_0z_\TA^4(t)}\int_0^\infty\id\id\!\!d\w\,\frac{\T{Im}r_{\,\,\id\T{p}}(\w)}{\w+\w_{10}}\\\label{eq:Fcpd2}
&=-\frac{3d^2\cos\theta}{8\pi^2\e_0z_\TA^4(t)}\int_0^\infty\id\id\!\!d\xi\,\frac{\w_{10}\;r_{\,\,\id\T{p}}(i\xi)}{\w^2_{10}+\xi^2}\,.
\end{align}
The cosine stems from the force's projection onto the direction of motion. If the atom's trajectory is chosen such that it moves towards the plane and hence $\cos\theta<0$, the projection of the Casimir-Polder force onto the velocity vector is positive, as it should be.

The leading-order-in-$v$ friction force according to Eq. \eqref{eq:FMarkov0} amounts to 
\begin{align}\nonumber
&F_\T{fr}(t)\simeq-\frac{3d^2}{8\pi^2\e_0z_\TA^4(t)}\int_0^\infty\id\id\!\!d\w\,\frac{\IM r_{\,\,\id\T{p}}(\w)}{\w+\Omega}\\\nonumber
&\times\li[\frac{v\Ga_1(1+\cos^2\!\theta)}{z_\TA(t)\,(\w+\Omega)^2}-\li(\de^vC^*_{01}+\de^vC_{10}\re)\cos\theta\re]\\\label{eq:FMarkov}
&\qq=-\frac{3d^2}{8\pi^2\e_0z_\TA^4(t)}\int_0^\infty\id\id\!\!d\xi\,\frac{\Omega\,r_{\,\,\id\T{p}}(i\xi)}{\Omega^2+\xi^2}\\\nonumber
&\times\li[\frac{v\Ga_1(\Omega^2\!-3\xi^2)(1\!+\cos^2\!\theta)}{z_\TA(t)\,(\Omega^2+\xi^2)^2}-\li(\de^vC^*_{01}+\de^vC_{10}\re)\cos\theta\re]\!.
\end{align}
Several remarks are in order at this point. First and foremost, note that the above expression is only correct up to order $d^4$, since the internal dynamics of the atom, entering the force, were only determined up to order $d^2$. Secondly, the \textit{implicit} $v$-dependence only contributes for non-parallel motion since it is weighted with a factor $\cos\theta$. Lastly note that, even though the leading-order in $v$ of the above force is linear, this linear order in velocity is of fourth order in the atomic dipole moment. If only $d^2$ terms are considered, the friction force vanishes to first order of $v$. This is because in this lowest order perturbation theory the rates and implicit velocity dependencies are set identically zero, i.e., $\Ga_1=\de^vC_{01}=0$. It can be shown that, on the $d^2$ level for parallel motion, the non-resonant force is strictly zero in all orders of atomic velocity. For any other direction of motion, its leading-order-in-$v$ force on level $d^2$ reads, 
\begin{align}\nonumber
F^{(2)}_\T{fr}(t)
&\simeq\frac{15\,d^2v^2}{16\pi^2\e_0z_\TA ^6(t)}\int_0^\infty\id\id\!\!d\xi\,\frac{\w_{10}(\w_{10}^2-3\xi^2)}{(\w_{10}^2+\xi^2)^3}\,r_{\,\,\id\T{p}}(i\xi)\\
&\qq \times (1+\cos^2\theta)\cos\theta\,.
\label{f2fricMark}
\end{align}
and is hence quadratic in the atomic velocity.

This provides us with a first answer to the initial question, whether quantum friction would be qualitatively different for non-parallel motion. Up to second order in coupling the answer is yes. Whereas the $d^2$ force on a parallely moving atom is exponentially suppressed as we saw above and as it was also previously shown in Ref.~\cite{Intravaia2015}, a non-parallely moving atom experiences an unsuppressed force which scales quadratic in velocity. This can be easily understood. On $d^2$ level the time-integral entering the force for parallel motion takes the form
\begin{align}
\RE\!\!\int_0^\infty\id\id\!\!d\tau\;e^{-i(\omega+\Omega-k^\pa v)\tau}=\de(\omega+\Omega-k^\pa v)\,, 
\end{align}
as illustrated before, this resonance condition enforces a constraint on the wave vector which in turn leads to the exponential suppression. For vertical motion, however, the time-integral entering the $d^2$ friction force reads
\begin{align}
-\IM\!\!\int_0^\infty\id\id\!\!d\tau\;e^{-[i(\omega+\Omega)+k^\pa v]\tau}=\frac{\omega+\Omega}{(\omega+\Omega)^2+(k^\pa v)^2}\,.
\end{align}
Instead of the sharp resonance condition in the co-moving frame, enforced by a Dirac-$\de$, we here encounter a Lorentzian whose width is given by the velocity. This Lorentzian is evidently quadratic in $v$. This very basic consideration also reveals how a finite linear order in $v$ comes about in the Markovian $d^4$ friction force. There, the same Lorentzian appears, however with a width now given by the sum of the atomic rate of spontaneous decay and the velocity. The latter hence broadens the Lorentzian peak in a linear manner. Note that this qualitative difference between parallel and vertical friction is intimately linked to the evanescent nature of near-field waves.

Lastly, note that from the expression \eqref{f2fricMark} the directionality of the force is not immediately evident. A more careful study of \eqref{f2fricMark} reveals that the integration of $(\w_{10}^2-3\xi^2)/(\w_{10}^2+\xi^2)^3$, from zero to infinite imaginary frequency, identically vanishes since the positive contribution for small imaginary frequencies exactly balances the negative contribution for large imaginary frequencies. With $r_p(i\xi)$ being strictly positive and monotonously decreasing this leads to overall positivity of the imaginary-frequency integral. For trajectories such that the atom moves towards the plane ($\cos\theta<0$), the projection of the frictional force onto the velocity is negative and counteracts the motion.
\end{subsection}
\end{section}

\begin{section}{Time-Dependent Perturbation Theory}

In this Section we will compute the quantum frictional force on a moving atom at constant velocity in arbitrarily direction motion using time-dependent perturbation theory. We will closely follow the approach used in Refs.~\cite{Barton2010,Intravaia2015}. To this end, the mathematical framework sketched in Section~\ref{sec:setup} is slightly modified. First of all, calculations are now performed in the interaction picture, rather than in the Heisenberg picture. Inspired by the 1s and 1p states of the hydrogen atom, the lowest quantum states of the atom are now taken to be the ground state $|g \rangle$ and three degenerate excited states written as $| \boldsymbol{\eta} \rangle$. The unit vector $\boldsymbol{\eta}$ is taken from a set $\{ \boldsymbol{\eta} \}$ forming an orthonormal and real basis. As before, the bare transition frequency between the levels is $\omega_{10}$ and the atom interacts with the electromagnetic field through its electric dipole momentum, $\hat{\di}(t)$. Its nonzero matrix elements in the interaction picture are $\bra{g} \hat{\di}(t) \ket{\boldsymbol{\eta}}=\boldsymbol{\eta} d e^{-i \omega_{10} t}$ and $\bra{\boldsymbol{\eta}} \hat{\di}(t)  \ket{g}=\boldsymbol{\eta} d e^{i \omega_{10} t}$. 

As mentioned in Section~\ref{sec:setup}, we assume that for $t<0$ the atom is static at a distance $z_0$ from the surface, and that for $t>0$ its distance from the surface varies as $z_A(t)=z_0 + v t \cos \theta$. 
As in the Markovian approach discussed before, we will also assume here that  $z_A(t)$ is always within the near-field zone, irrespective of whether the atom is moving towards or away from the surface. Hence the electric field operator  can be written as \cite{Barton2010,Intravaia2015}
\begin{align}
\hat{\E}(\R_\TA)
&= \int\!\! d^2\K^\pa\!\!\int_0^{\infty}\id\id\!\! d\omega\; i \K \;\hat{a}_{\K^\pa \omega} \psi_{\K^\pa \omega} \\\nonumber
&\qq\times e^{i \K^\pa \cdot \R_\TA^\pa(t)-i\omega t - k^\pa z_\TA(t)} + \text{h.c.}
\end{align}
Here, $\hat{a}_{\K^\pa \omega}$, $\hat{a}^{\dagger}_{\K^\pa \omega}$ are bosonic annihilation/creation operators (which roughly correspond to the 2D Fourier transform of the $\f^\dag_\T{e}(\R,\w)$ in Section~\ref{sec:setup}), and $\psi_{\K^\pa \omega}$ are complex plasmon amplitudes whose modulus squared is $|\psi_{\K^\pa \omega}|^2= (\hbar/(8\pi^3\e_0 k^\pa)) \text{Im} r_{\rm p}(\omega)$. Note that they differ by a factor $\sqrt{4\pi\e_0}$ with respect to the aforementioned references. This is rooted in the use of different unit systems. Here we employ SI units whereas in those references Gaussian units were used.

We assume that the initial state of the system is the atom in its ground state and no photons, i.e. $\ket{\psi (0)}= \ket{g;\text{vac}}$. As in \cite{Intravaia2015}, we express the joint atom+field state in a perturbative expansion in the coupling constant $d$. To third order, it is given by 
\begin{align}
\ket{\psi (t)}
&\simeq\left(1+c_0^{(2)}(t)\right) \ket{g;\text{vac}} \\\nonumber
&\qq+ \sum_{\boldsymbol{\eta}} \int\id d^3 \kappa \left(c_1^{(1)}(t) + c_1^{(3)}(t)\right) \ket{\boldsymbol{\eta}; \kappa}\\\nonumber
&\qq+ \tfrac{1}{2}\int\id d^3 \kappa_1\!\! \int\id d^3 \kappa_2 \;c_2^{(2)}(t) \ket{g;\kappa_1,\kappa_2}\,,
\end{align}
where $c_n^{(p)}(t)$ denotes the transition amplitudes for states with $n$ photons in the $p$th perturbative order and can be obtained by using the standard techniques of perturbation theory. We need to compute the state to third order in order to evaluate the force to fourth order in the coupling. Above we have used the compact notation $\kappa=\{\K^\pa,\omega\}$, and the integrals $\int d^3 \kappa =\int d^2\K^\pa \int_0^{\infty} d\omega$.

The expectation value in the state $\ket{\psi (t)}$ of the force operator along the direction of motion, $\hat{F}_v = \V \cdot \hat{\F}/v = \sin \theta \hat{F}_x +\cos \theta \hat{F}_z$, is given by
\begin{align}\label{eq:fza}
&F_v(t)
= 2 \text{Re}\sum_{\boldsymbol{\eta}} \left\{\int\!\!d^3\kappa \bra{g;\text{vac}} \hat{F}_v
\ket{\boldsymbol{\eta}; \kappa} \right.\\\nonumber
&\qq\qq\qq\times \left[c_1^{(1)}(t)+c_0^{(2)*}(t) c_1^{(1)}(t)+c_1^{(3)}(t)\right] \nonumber \\\nonumber
&+ \left.\tfrac{1}{2}\int\!\!d^3\kappa\,d^3\kappa_1\,d^3\kappa_2\, 
\bra{\boldsymbol{\eta}; \kappa} \hat{F}_v \ket{g;\kappa_1,\kappa_2}c_1^{(1)*}(t) c_2^{(2)}(t) \right\} ,
\end{align}
valid to fourth order in the coupling. The relevant matrix elements of the interaction Hamiltonian are 
\begin{align}
& \bra{g;\text{vac}}\hat{H}_{\rm AF} \ket{\boldsymbol{\eta};\kappa}=i d (\boldsymbol{\eta} \cdot \K) \psi_{\kappa}\\\nonumber
&\qq\times e^{-i(\omega_{10}+\omega)t+i \K^\pa \cdot \R_\TA^\pa(t)- k^\pa z_\TA(t)}\,,\\
&\bra{\boldsymbol{\eta};\kappa} \hat{H}_{\rm AF} \ket{g;\kappa_1,\kappa_2} =i d  (\boldsymbol{\eta} \cdot \K_1) \psi_{\kappa_1}\\\nonumber
&\qq\times e^{i(\omega_{10}-\omega_1)t+i\bm{\K^\pa_1}\cdot \R_\TA^\pa(t)- k^\pa_1 z_\TA(t)} \delta^3(\kappa-\kappa_2)\\\nonumber
&\qq+ (1 \leftrightarrow 2)\,, \nonumber \\
& \bra{\boldsymbol{\eta};\text{vac}} \hat{H}_{\rm AF} \ket{g;\kappa}= i d (\boldsymbol{\eta} \cdot \K) \psi_{\kappa}\\\nonumber
&\qq\times e^{i (\omega_{10}-\omega)t+i \K^\pa \cdot \R_\TA^\pa(t)- k^\pa z_\TA(t)}\,, 
\end{align}
where $\delta^3(\kappa-\kappa_1) = \delta^2(\K^\pa-\K^\pa_1) \delta(\omega-\omega_1)$. The relevant matrix elements of the force operator are easily computed. We obtain
\begin{align}
&\bra{g;\text{vac}} \hat{F}_v \ket{\boldsymbol{\eta}; \kappa} = - i d k^\pa (\boldsymbol{\eta} \cdot \K) f_{\phi\theta} \psi_{\kappa}\\\nonumber
&\qq\times e^{-i (\omega_{10} + \omega)t + i \K^\pa \cdot \R_\TA^\pa(t)-k^\pa z_\TA(t)}\,,\\
&\bra{\boldsymbol{\eta}; \kappa} \hat{F}_v \ket{g;\kappa_1,\kappa_2} =- i d k^\pa_1 (\boldsymbol{\eta}  \cdot \K_1) f_{\phi_1 \theta} \psi_{\kappa_1}\\\nonumber
&\qq\times e^{i (\omega_{10} - \omega_1)t + i \K^\pa_1\cdot \R_\TA^\pa(t)-k^\pa_1 z_\TA(t)} \delta^3(\kappa-\kappa_2)\\\nonumber
&\qq+ (1 \leftrightarrow 2)\,,
\end{align}
where $f_{\phi\theta} =-\cos\theta + i \sin \theta \cos\phi$. 

\begin{subsection}{Internal Atomic Dynamics}

We now compute the relevant transition amplitudes $c_n^{(p)}(t)$ necessary to evaluate the force to fourth order in perturbation theory. The coefficient $c_1^{(1)}(t)$ is given by
\begin{align}
\label{c11}
c_1^{(1)}(t)
&=- \tfrac{i}{\hbar} \int_0^t  dt' \bra{\boldsymbol{\eta};\kappa} \hat{H}_{\rm AF}(t') \ket{g;\text{vac}}\\\nonumber
&= \frac{i d (\boldsymbol{\eta} \cdot \K)^* \psi^*_{\kappa}}{\hbar (\omega_{10} +\omega')}\;
e^{-i \K^\pa \cdot \R_0- k^\pa z_0} 
\left[ e^{i(\omega_{10}+\omega')t} -1 \right]\,,
\end{align}
where we have defined the complex frequency
\begin{align}
\omega'=\omega - v k^\pa \cos \phi \sin \theta + i v k^\pa \cos \theta\,,
\end{align}
and $\R_0=(x_0,y_0)$. 

The coefficient $c_2^{(2)}(t)$ is given by
\begin{align}
\label{c22}
&c_2^{(2)}(t)
=- \tfrac{i}{\hbar}\sum_{\boldsymbol{\eta}}\!\!\int\!\!d^3\kappa\!\!\int_{0}^t dt'c_1^{(1)}(t') \bra{g;\kappa_1 \kappa_2 }\hat{H}_{\rm AF}(t')\ket{\boldsymbol{\eta};\kappa}\\\nonumber
&=\!- \frac{d^2 (\K_1 \cdot \K_2)^* \psi_{\kappa_1}^* \psi_{\kappa_2}^* }{\hbar^2}
e^{-i (\K^\pa_1+\K^\pa_2) \cdot \R_0 - (k^\pa_1+k^\pa_2) z_0}  \\ \nonumber
& \times  
\left[ 
\frac{e^{i (\omega'_1+\omega'_2) t} -1}{\omega_1'+\omega_2'}  
\left( \frac{1}{\omega_{10}+\omega'_1} + \frac{1}{\omega_{10}+\omega'_2} \right)
\right. \\ \nonumber
& 
\left.
+ \frac{e^{-i (\omega_{10}-\omega'_1)t}-1}{(\omega_{10}-\omega'_1) (\omega_{10}+\omega'_2)}
+ \frac{e^{-i (\omega_{10}-\omega'_2)t}-1}{(\omega_{10}-\omega'_2) (\omega_{10}+\omega'_1)}
\right], 
\end{align}
with the separately shifted frequencies
\begin{align}
\omega'_j=\omega_j-v k^\pa_j\cos\phi_1 \sin\theta + i v k^\pa_j \cos\theta\,.
\end{align}

The coefficient $c_0^{(2)}(t)$ is given by
\begin{align}\nonumber
c_0^{(2)}(t)
&=-\tfrac{i}{\hbar}\sum_{\boldsymbol{\eta}}\!\!\int\!\!d^3\kappa\!\!\int_0^t dt'c_1^{(1)}(t')\bra{g;\text{vac}}\hat{H}_{\rm AF}(t') \ket{\boldsymbol{\eta};\kappa}\\\nonumber
&=-\tfrac{i d^2}{4\pi^3\hbar\e_0}\!\!\int_0^{\infty}\id\id\!\!d\omega \text{Im}r_{\rm p}(\omega)
\!\!\int\!\!d^2\K^\pa 
\frac{k^\pa e^{-2 k^\pa z_0}}{\omega_{10} +\omega'} 
\\\label{c02}
& \times 
\left[
\frac{e^{-2 k^\pa v \cos\theta t}-1}{-2 k^\pa v \cos\theta} -
\frac{e^{-i(\omega_{10} + \omega')t -2 k^\pa v \cos\theta t}-1}{-i(\omega_{10} + \omega') -2 k^\pa v \cos\theta}
\right]
\end{align}
This coefficient involves the energy shift of the state $\ket{g;{\rm vac}}$ and the rate for the process $\ket{g;{\rm vac}} \rightarrow \ket{\boldsymbol{\eta};\kappa}$. In the limit $k^\pa v t \ll 1$ (small times or small velocities), the first term within the square brackets grows as $t$, while the second one is subleading in time (it is a sum of an modulated oscillatory function plus a time-independent term). We can then approximate $c_0^{(2)}(t)$ as
\begin{align}
c_0^{(2)}(t) \simeq - \frac{i t}{\hbar} \delta E_g -  t \frac{\Gamma_g}{2} ,
\end{align}
and hence $1+ c_0^{(2)}(t) \simeq \exp[- \frac{i t}{\hbar} \delta E_g -  t \frac{\Gamma_g}{2}]$,
where $\delta E_g$ is the energy shift and $\Gamma_g$ is the rate. 
Performing a further expansion of the integrand in (\ref{c02}) in powers of $k^\pa v$ and carrying out the 
momentum integration, we obtain 
\begin{align}
\delta E_g 
&\simeq-\tfrac{d^2}{8\pi^2 \e_0 z_0^3} 
\int_0^\infty\id\id\!\!d\w\;\tfrac{\IM r_p(\w)}{\w+\w_{10}}
\left[
1-\tfrac{3}{4}\tfrac{v^2(1+3\cos2\theta)}
{(\w+\w_{10})^2z_0^2}
\right],
\label{Eg}
\end{align}
and
\begin{align}
\Gamma_g
& \simeq -\tfrac{3 d^2v\cos\theta}{8\pi^2\hb\e_0z_0^4}
\int_0^\infty\id\id\!\!d\w\;\tfrac{\IM r_p(\w)}{(\w+\w_{10})^2} .
\label{Gg}
\end{align}
These equations for the shift and rate coincide with Eqs. (\ref{eq:Ga01nres}) and (\ref{eq:dw01}) obtained in the Markovian approach, with the slight difference that in the latter the instantaneous height $z_\TA(t)$ rather than the initial height $z_0$ appears. We note that the rate (\ref{Eg}) vanishes for parallel motion, consistent
with the exponentially small rate found in the perturbative approach for parallel motion studied in Refs.~\cite{Barton2010,Intravaia2015}. 

Finally, we compute the $c_1^{(3)}(t)$ coefficient, which we express as a sum of two contributions $c_1^{(3)}(t) = c_{1,0}^{(3)}(t) + c_{1,2}^{(3)}(t)$. The subscript 0 in the first term denotes contributions from the vacuum, and the subscript 2 in the second term denotes those from the two-photon sector. They are respectively given by
\begin{align}\label{c13-0}
&c_{1,0}^{(3)}(t)=-\tfrac{i}{\hbar}\!\!\int_0^t dt'c_0^{(2)}(t')\bra{\boldsymbol{\eta};\kappa} \hat{H}_{\rm AF}(t') \ket{g;\text{vac}}\\\nonumber
&\q=\tfrac{d (\boldsymbol{\eta} \cdot \K)^* \psi_{\kappa}^*}{\hbar}\!\!\int_0^t dt' 
e^{i(\omega_{10}+\omega)t'-i \K^\pa \cdot \R^\pa_\TA(t')-k^\pa z_\TA(t')} \\\nonumber
&\qq\qq\times t' \left( \tfrac{\Gamma_g}{2} + \tfrac{i \delta E_g}{\hbar} \right)
\end{align}
and 
\begin{align}\label{c13-2}
&c_{1,2}^{(3)}(t)= \tfrac{i}{2\hbar}\!\!\int\!\!d^3\kappa_1 d^3\kappa_2\!\!\int_0^t dt'c_2^{(2)}(t')\bra{\boldsymbol{\eta};\kappa} \hat{H}_{\rm AF}(t')\ket{g; \kappa_1 \kappa_2}\\\nonumber
&= \frac{i d^3}{\hbar^3} \psi_{\kappa}^*  
e^{-i \K^\pa \cdot \R_0}
\int\!\!d^3\kappa_1 
\frac{(\boldsymbol{\eta} \cdot \K_1) (\K_1 \cdot \K)^* |\psi_{\kappa_1}|^2 e^{-2k^\pa_1z_0}}
{\omega'_1+\omega'}  \\ \nonumber
&
\times
\left\{
\left[ \frac{1}{\omega_{10}+\omega'_1} + \frac{1}{\omega_{10}+\omega'} 
\right]
\right. 
\\ \nonumber
&
\left[ 
\frac{e^{i(\omega_{10}+\omega'+2 i v k^\pa \cos\theta)t}-1}{\omega_{10}+\omega'+2 i v k^\pa \cos\theta} 
-
\frac{e^{i(\omega_{10}-\omega'_1+2 i v k^\pa \cos\theta)t}-1}{\omega_{10}-\omega'_1+2 i v k^\pa \cos\theta}
\right] \\ \nonumber
& +
\frac{1}{(\omega_{10}+\omega') (\omega_{10}-\omega'_1)}
\\ \nonumber
&
\left.
\left[ 
\frac{e^{-2 v k^\pa \cos\theta t}-1}{2 i v k^\pa \cos\theta} 
-
\frac{e^{i(\omega_{10}-\omega'_1+2 i v k^\pa \cos\theta)t}-1}{\omega_{10}-\omega'_1+2 i v k^\pa \cos\theta}
\right] 
\right\}.
\end{align}

\end{subsection}

\begin{subsection}{Casimir-Polder and Friction Force: 2nd Order}

The first non-vanishing contribution to the force is second order in the coupling, and is given by the term in Eq.(\ref{eq:fza}) containing only the $c_1^{(1)}(t)$ coefficient. We obtain
\begin{align} \label{f2}
&F^{(2)}_v(t)=2 \text{Re} \sum_{\boldsymbol{\eta}} \int\!\!d^3\kappa \bra{g;\text{vac}} \hat{F}_v  \ket{\boldsymbol{\eta}; \kappa}c_1^{(1)}(t)\\\nonumber
&=\frac{d^2}{2\pi^3\e_0}\int_0^{\infty}\id\id\!\!d\omega\!\!\int\!\!d^2\K^\pa k^{\pa2} e^{-2 k^\pa z_\TA(t)}\text{Im}r_{\rm p}(\omega) \\\nonumber
& \times {\rm Re}\!\left[ \tfrac{f_{\phi\theta}}{\omega_{10}+\omega'}  \left(1-e^{-i (\omega_{10}+\omega-v k^\pa \cos\phi \sin\theta)t} \right)
\right].
\end{align}
The second term within the square brackets leads to a modulated oscillatory contribution to the force, and averages out to zero after time averaging. 
The first term, on the other hand, gives a non-vanishing contribution, and its explicit form can be evaluated in the low-velocity limit. To this end, it is 
convenient to introduce the dimensionless variables $s=k^\pa z_\TA(t)$ and $y=v/[z_\TA(t) (\omega_{10}+\omega)]$. Then the force is re-written as
\begin{align}\label{f2tot}
F^{(2)}_v(t)
&=-\frac{d^2}{2\pi^3\e_0}\frac{\cos \theta}{z_\TA^4(t)}\int_0^\infty\id\id\!\!d\omega\;\frac{\text{Im} r_{\rm p}(\omega)}{\omega_{10}+\omega}\\\nonumber
&\q\times\!\!\int_0^{2\pi}\id\id\!\!d\phi\!\int_0^\infty\id\id\!\!ds\,\frac{s^3 e^{-2s} (1-2 y s \cos \phi \sin \theta)}{(1-y s \cos\phi \sin\theta)^2+(y s \cos \theta)^2}.
\end{align}
In the adiabatic regime $y\ll 1$ in which the characteristic frequency of the motion $v/z_\TA(t)$ is much smaller than the atom's transition frequency $\omega_{10}$, we can express the force as the sum of two contributions  $F_v^{(2)}(t) = F_{\rm CP}^{(2)}(t) + F_{\text{fr}}^{(2)}(t)$, where
\begin{align}\nonumber
F_{\rm CP}^{(2)}(t)
&=-\frac{3 d^2}{8\pi^2\e_0} \frac{\cos \theta}{z^4_\TA(t)} \int_0^\infty\id\id\!\!d\omega\;\frac{\text{Im} r_{\rm p}(\omega)}{ \omega_{10}+\omega}\\\label{fcp2pert}
&=-\frac{3 d^2}{8\pi^2\e_0} \frac{\cos \theta}{z^4_\TA(t)} \int_0^\infty\id\id\!\!d\xi\;\frac{\omega_{10}}{\omega^2_{10}+\xi^2}\;r_{\rm p}(i\xi) 
\end{align}
is the projection of the standard Casimir-Polder force along the direction of the motion, evaluated at the instantaneous position of the atom. In the last step we have Wick-rotated to imaginary frequencies $\omega \rightarrow i \xi$. This perturbative result perfectly coincides with the expression \eqref{eq:Fcpd2} obtained via the Markovian approach.  Note that for parallel motion ($\theta = \pi/2$) this projection is zero, as expected, since the drag force is orthogonal to the Casimir-Polder force. Also note that for vertical motion  ($\theta=0, \pi$), $F_{\rm CP}^{(2)}(t)$ changes sign, which is simply due to the change of sign of the velocity vector. The other force  term $F_{\text{fr}}^{(2)}(t)$ is given by
\begin{align}\nonumber
F_{\text{fr}}^{(2)}(t)
&=\frac{15 d^2 v^2 \cos \theta (1+\cos^2\theta)}{16\pi^2\e_0 z^6_\TA(t)}\int_0^\infty\id\id\!\!d\omega\,\frac{\text{Im} r_{\rm p}(\omega) }{(\omega_{10}+\omega)^3}\\\nonumber
&=\frac{15 d^2 v^2 }{16\pi^2\e_0 z^6_\TA(t)} \int_0^{\infty}\id\id\!\!d\xi\,\frac{\omega_{10} (\omega_{10}^2-3 \xi^2)}{ (\omega_{10}^2+\xi^2)^3} r_{\rm p}(i \xi)\\\label{fric2pert}
&\qq\times(1+\cos^2\theta)\cos \theta\,.
\end{align}
This equation coincides with Eq.~(\ref{f2fricMark}) obtained via the Markovian approach. Note that $F_{\text{fr}}^{(2)}(t)=0$ for parallel motion, in agreement with previous works in the literature that showed that quantum friction to second order in the coupling is vanishingly small (see, for example Refs. \cite{Barton2010,Intravaia2015}). 
\end{subsection}

\begin{subsection}{Casimir Polder and Friction Force: 4th Order, via Vacuum}

We now compute the next order of the force, which is fourth-order in the coupling. To this end, we follow an approach similar to the one described in Appendix C of Ref.~\cite{Intravaia2015}. We first consider the part involving the mixed amplitude $c_0^{(2)*}(t) c_1^{(1)}(t)$ in Eq.~(\ref{eq:fza}) and the part of $c_1^{(3)}(t)$ going via vacuum, Eq.~(\ref{c13-0}). Integrating by parts and using Eq.~\eqref{c11}, 
$c_{1,0}^{(3)}(t)$ can be written as
\begin{align} \label{inter1}
& c_{1,0}^{(3)}(t)= c^{(2)}_0(t) \; c^{(1)}_1(t) - \frac{i d (\boldsymbol{\eta} \cdot \K)^* \psi^*_{\kappa}}{\hbar (\omega_{10} +\omega')}\;
c^{(2)}_0(t)
e^{-i \K^\pa \cdot \R_0- k^\pa z_0} \\\nonumber
& +    \left(- \frac{1}{\hbar} \delta E_g + i \frac{\Gamma_g}{2}\right)
\frac{d (\boldsymbol{\eta} \cdot \K)^* \psi^*_{\kappa}}{\hbar (\omega_{10} +\omega')^2}\;
e^{-i \K^\pa \cdot \R_0- k^\pa z_0} \\\nonumber
& \times [e^{i(\omega_{10} +\omega')t}-1].
\end{align}
Combining the first term above with $c_0^{(2)*}(t) c_1^{(1)}(t)$ results in $-\Gamma_g t c_1^{(1)}(t)$, and then the corresponding fourth-order force is 
\begin{equation}
F^{(4)}_{v,0}(t)= -\Gamma_g t F^{(2)}_v(t),
\label{f40}
\end{equation}
which, when combined with the force at second order $F^{(2)}_v  (t)$, represents a loss of probability in the ground state. 
The subscript 0 in $F^{(4)}_{v,0}(t) $ denotes contributions from the vacuum. The contribution to the force of the second term in \eqref{inter1} is
\begin{align} \label{van1}
& -\frac{d^2}{2 \pi^3  \epsilon_0} \int_0^{\infty} d\omega \text{Im} r_{\rm p}(\omega) \int d^2k^\pa (k^\pa)^2 e^{-2 k^\pa z_0 - k^\pa v t \cos\theta} \\\nonumber
& \times \text{Re} \left[ \frac{f_{\phi \theta}}{\omega_{10}+\omega'} 
\left(\frac{i t}{\hbar} \delta E_g + t \frac{\Gamma_g}{2}\right)
e^{-i (\omega_{10} + \omega - k^\pa v \cos\phi \sin \theta)t} \right].
\end{align}
This is a modulated oscillatory function of time, and vanishes after time-averaging. The contribution to the force of the third term in \eqref{inter1} is
\begin{align} \label{van2}
& \frac{d^2}{2 \pi^3 \hbar \epsilon_0} \int_0^{\infty} d\omega \text{Im} r_{\rm p}(\omega) \int d^2k^\pa (k^\pa)^2 e^{-2 k^\pa z_\TA(t)} \\\nonumber
& \times \text{Re} \left\{ 
\left( - \frac{1}{\hbar} \delta E_g + i \frac{\Gamma_g}{2} \right)\frac{f_{\phi \theta}}{(\omega_{10}+\omega')^2} \right\} + \text{m.o.t.} ,
\end{align}
where "m.o.t." denotes modulated oscillatory terms -- similar to the ones appearing in \eqref{van1} -- that
vanish after time-averaging. In order to evaluate \eqref{van2}, we re-write the second line in terms of the dimensionless variables $s=k^\pa z_\TA(t)$ and $y=v/[z_\TA(t) (\omega_{10}+\omega)]$ introduced above, and perform an expansion in powers of velocity. To order $y^0$, we obtain a $d^4$-correction to the Casimir-Polder force \eqref{fcp2pert},
\begin{align}\label{FCP4}
F_{\rm CP}^{(4)}(t)
&=\frac{3 d^2 \delta E^{(0)}_g}{8\pi^2 \hbar \e_0} \frac{\cos \theta}{z^4_\TA(t)} \int_0^\infty\id\id\!\!d\omega\;\frac{\text{Im} r_{\rm p}(\omega)}{(\omega_{10}+\omega)^2}.
\end{align}
where $\delta E^{(0)}_g$ is the velocity-independent term of the shift $\delta E_g$ defined in \eqref{Eg}. 
Note that this correction identically vanishes for parallel motion. To order $y^1$, we obtain a 
$d^4$-correction to the friction force \eqref{fric2pert},
\begin{align}\label{Ffr4}
F_{\rm fr}^{(4)}(t)
&=- \frac{3 d^2 \Gamma_g v (1+\cos^2 \theta)}{8\pi^2 \hbar \e_0 z^5_\TA(t)} \int_0^\infty\id\id\!\!d\omega\;\frac{\text{Im} r_{\rm p}(\omega)}{(\omega_{10}+\omega)^3}.
\end{align}
Recalling the definition of the rate $\Gamma_g$ in \eqref{Gg}, we conclude that $F_{\rm fr}^{(4)}(t)$ goes as $v^2$ and vanishes for parallel motion.

\end{subsection}

\begin{subsection}{Casimir Polder and Friction Force: 4th Order, via Two Photons}\label{sec:v3}

The final piece for the fourth order force contains two contributions. One arises from the part of $c_1^{(3)}(t)$ that involves the two-photon sector, i.e.,  $c_{1,2}^{(3)}(t)$ in Eq.(\ref{c13-2}), and another one from the coherence between the one- and two-photon sectors, last term in Eq.(\ref{eq:fza}). We denote them by $F^{(4)[03]}_{v,2}(t)$ and $F^{(4)[12]}_{v,2}(t)$, respectively, and we compute them separately. The subscript 2 in $F^{(4)}_{v,2}(t) $ denotes contributions from the two-photon sector. 
They read
\begin{align}\nonumber
&F^{(4)[03]}_{v,2}(t) = \frac{d^4}{4 \pi \epsilon_0 \hbar^3}\;{\rm Re}\!\!\int\!\!d^3\kappa_1 d^3\kappa_2| \K_1 \cdot \K_2 |^2|\psi_{\kappa_1}|^2  |\psi_{\kappa_2}|^2\\\nonumber
&\q\times  e^{-2(k^\pa_1+k^\pa_2) z_\TA(t)} \left[ \tfrac{k^\pa_1 f_{\phi_1\theta}}{\omega_{10}+\omega'_1 + 2 i v k^\pa_2 \cos\theta} + (1 \leftrightarrow 2) \right]\\
&\q\times \tfrac{1}{\omega'_1 + \omega'_2}\left[ \tfrac{1}{\omega_{10}+\omega'_2} + (1 \leftrightarrow 2) \right] +
\text{m.o.t.}, 
\end{align}
and
\begin{align}\nonumber
&F^{(4)[12]}_{v,2}(t) = \frac{d^4}{4 \pi \epsilon_0 \hbar^3}\;{\rm Re}\!\!\int d^3\kappa_1 d^3\kappa_2 | \K_1 \cdot \K_2 |^2|\psi_{\kappa_1}|^2  |\psi_{\kappa_2}|^2\\\nonumber
&\q\times  e^{-2(k^\pa_1+k^\pa_2) z_\TA(t)} \left[ \tfrac{k^\pa_1 f_{\phi_1\theta}}{\omega_{10}+\omega'_2} + (1 \leftrightarrow 2) \right]\\
&\q\times \tfrac{1}{\omega'_1 + \omega'_2}\left[ \tfrac{1}{\omega_{10}+\omega'_2} + (1 \leftrightarrow 2) \right] 
+ \text{m.o.t.}.
\end{align}
The modulated oscillatory terms (m.o.t.) contributions vanish after time-averaging, and we will discard them in the following. Note that the two equations above have the same structure, except for the bracket in the second line of each of them. Defining their sum as $\Sigma^{(4)}(t) = F^{(4)[03]}_{v,2}(t) + F^{(4)[12]}_{v,2}(t)$, we obtain
\begin{align}\label{Sigma4}
&\Sigma^{(4)}(t)  = \frac{d^4}{4 \pi \epsilon_0\hbar^3}\;{\rm Re}\!\!\int\!\!d^3\kappa_1 d^3\kappa_2| \K_1 \cdot \K_2 |^2|\psi_{\kappa_1}|^2  |\psi_{\kappa_2}|^2\\\nonumber
&\q\times \frac{ e^{-2(k^\pa_1+k^\pa_2) z_\TA(t)} }{\omega'_1 + \omega'_2}\left[ \frac{1}{\omega_{10}+\omega'_2} + (1 \leftrightarrow 2) \right]\\\nonumber
&\q\times\left[ k^\pa_1 f_{\phi_1\theta} \left(\tfrac{1}{\omega_{10}+\omega'_1 + 2 i v k^\pa_2 \cos\theta} + \tfrac{1}{\omega_{10}+\omega'_2}\right)+ (1 \leftrightarrow 2) \right] .
\end{align}
In order to evaluate Eq. (\ref{Sigma4}), we proceed as in the previous subsections, and define variables $y_1=v/[z_\TA(t) (\omega_{10}+\omega_1)]$ and $y_2=v/[z_\TA(t) (\omega_{10}+\omega_2)]$, and perform an expansion in powers of $y_1$ and $y_2$. The calculations are quite cumbersome, and here we only report the main results. To lowest order in velocity (i.e. terms proportional to $y_1^0 y_2^0$), we obtain
\begin{align}\nonumber
\Sigma^{(4)}_0(t)
&=-\frac{3 d^4}{128 \pi^3 \hbar\e_0} \frac{\cos\theta}{z_\TA^7(t)} \int_0^{\infty}\id\id\!\!d\omega_1 d\omega_2\text{Im} r_{\rm p}(\omega_1) \text{Im} r_{\rm p}(\omega_2)\\
&\q\times \frac{(2 \omega_{10}+\omega_1+\omega_2)^2}{(\omega_1 + \omega_2) (\omega_{10}+\omega_1)^2 (\omega_{10}+\omega_2)^2}\,.
\end{align}
The subscript $0$ in $\Sigma^{(4)}_0(t)$ denotes zero-order in velocity. Hence, $\Sigma^{(4)}_0(t)$ is a correction to the Casimir-Polder force \eqref{fcp2pert}, coming from processes concerning the emission or absorption of two photons. The linear-in-velocity force (arising from terms proportional to $y_1^1 y_2^0$ and $y_1^0 y_2^1$) vanishes identically, i.e. $\Sigma^{(4)}_1(t) =0$. The next non-vanishing order is quadratic in velocity (it arises from terms proportional to $y_2^1 y_2^0$, $y_1^0 y_2^2$, and $y_1^1 y_2^1$), and the resulting force is $\Sigma^{(4)}_2(t) \propto v^2 \cos\theta / z_\TA^9(t)$ (the prefactor is a complicated integral over $\omega_1$ and $\omega_2$, and we do not report it here). Note that for parallel motion ($\theta=\pi/2$) both $\Sigma^{(4)}_0(t)$ and $\Sigma^{(4)}_2(t)$ are zero, and the first non-vanishing term is proportional to $v^3$, in agreement with the result found in Ref.~\cite{Intravaia2015}.
\end{subsection}
\end{section}

\begin{section}{Conclusions}
	
\begin{table}
\begin{tabular}{|m{2cm}|m{3cm}|m{3cm}|}
\hline       \rule[-2ex]{0pt}{5.5ex}                       & \q Markov                                                                          & \qq Perturbation                                   \\ 
\hline\hline \rule[-2ex]{0pt}{5.5ex} \q$\pa$ shift         & \qq$\left(\!\frac{v}{\w_{10}z_\TA}\!\right)^{\!2}$                                 & \qq$\left(\!\frac{v}{\w_{10}z_\TA}\!\right)^{\!2}$ \\ 
\hline       \rule[-2ex]{0pt}{5.5ex} \q$\perp$ shift       & \qq$\left(\!\frac{v}{\w_{10}z_\TA}\!\right)^{\!2}$                                 & \qq$\left(\!\frac{v}{\w_{10}z_\TA}\!\right)^{\!2}$ \\ 
\hline\hline \rule[-2ex]{0pt}{5.5ex} \q$\pa$ rate          & \qq exp. small                                                                     & \qq 0                                              \\ 
\hline       \rule[-2ex]{0pt}{5.5ex} \q$\perp$ rate        & \qq$\frac{v}{\w_{10}z_\TA}$                                                        & \qq$\frac{v}{\w_{10}z_\TA}$                        \\ 
\hline\hline \rule[-2ex]{0pt}{5.5ex} \q$\pa$ force $d^2$   & \qq exp. small                                                                     & \qq 0                                              \\ 
\hline       \rule[-2ex]{0pt}{5.5ex} \q$\perp$ force $d^2$ & \qq$\left(\!\frac{v}{\w_{10}z_\TA}\!\right)^{\!2}$                                 & \qq$\left(\!\frac{v}{\w_{10}z_\TA}\!\right)^{\!2}$ \\ 
\hline\hline \rule[-2ex]{0pt}{5.5ex} \q$\pa$ force $d^4$   & \qq$\left(\frac{v}{\omega_{10}z_\TA}\right)\left(\frac{\Ga_1}{\omega_{10}}\right)$ & \qq$\left(\!\frac{v}{\w_{10}z_\TA}\!\right)^{\!3}$ \\ 
\hline       \rule[-2ex]{0pt}{5.5ex} \q$\perp$ force $d^4$ & \qq$\left(\frac{v}{\omega_{10}z_\TA}\right)\left(\frac{\Ga_1}{\omega_{10}}\right)$ & \qq$\left(\!\frac{v}{\w_{10}z_\TA}\!\right)^{\!2}$ \\ 
\hline 
\end{tabular}
\caption{Comparison of the motion-induced corrections to both the internal dynamics of a ground-state atom moving in front of a macroscopic body and the Casimir-Polder force that acts upon that atom. The table gives the scaling of shifts, rates and forces with increasing atomic speed $v$, bare atomic transition frequency $\omega_{10}$, atom-surface separation $z_\TA$ and atomic rate of spontaneous decay $\Ga_1$. Here perpendicular ($\perp$) refers to an atom moving towards the body. The results obtained via Markovian quantum master equations (``Markov'') and time-dependent perturbation theory (``Perturbation'') agree for the level shift and rate, as well as the $d^2$ level friction force, but differ for the $d^4$ friction force.}
\label{tab:results}
\end{table} 

We summarize our results for the motion-induced corrections to both the internal dynamics -- energy level and rate of transition -- of a ground-state atom and the Casimir-Polder force that acts upon that very atom in Table~\ref{tab:results}. In this table, non-parallel motion is represented by its extreme, i.e., perfectly vertical motion of the atom towards the surface. The results for such motion are contrasted with the ones for the parallel scenario. In both cases, the Markovian approach and time-dependent perturbation theory agree concerning leading order dynamical corrections to level shifts (compare Eqs.~\eqref{eq:dw01} and \eqref{Eg}) and decay rates (compare Eqs.~\eqref{eq:Ga01res}, \eqref{eq:Ga01nres} and \eqref{Gg}). Note that the Markovian approach predicts an exponentially small parallel rate, while this rate is identically zero in the perturbative approach. Regarding the quantum frictional force, both approaches agree to second order in the atom-field coupling. For example, they both predict a vanishing force for parallel motion, and a $v^2$ scaling for vertical motion (compare Eqs.~\eqref{f2fricMark} and \eqref{fric2pert}). In contrast, the two approaches differ to fourth order in the coupling, irrespective of the direction of motion. For example, within the Markovian approach the force is always linear in $v$ for all directions, while according to time-dependent perturbation theory the friction force undergoes a qualitative change from a $v^3$ behavior for parallel motion and a $v^2$ scaling for vertical motion instead.

As explained in more detail in the introduction, at this point in time we cannot decide whether any of the two results is flawed -- and if so, for which reason -- or whether they merely apply to different temporal regimes. Both the Markovian and the perturbative approach rely on approximations which restrict their respective applicability. The Markov approximation presupposes exponential decay of both one- and two-point expectation values on all timescales, which in turn implies perfectly Lorentzian power spectra of all observables. This restricts the realm of validity of the such obtained results in twofold manner: firstly, the Markovian results only apply to a spectral range near atomic and surface plasmon resonances and secondly, as a consequence, only apply to timescales smaller than those where algebraic decay sets in. More interesting when comparing to perturbation theory, however, is the fact that the Markovian results also do not apply to timescales shorter than the electromagnetic field's auto-correlation time. It is on this timescale that memory effects lead to transient behavior which is not resolved when applying the Markov approximation. These short times -- fractions of the atomic excited state's lifetime --  though, are the domain of time-dependent perturbation theory. Therefore, one need not be surprised if Markovian and perturbative results for both the internal dynamics of the atom and the quantum friction force do not agree. However, it would be desirable to map out the transition from one regime to the other or verify the respective results by means of a third theoretical method or experiment.

Finally, be reminded that the facilitation of quantum friction force measurements -- which as of now are far out of reach -- was the original motivation for this work. Inspired by the findings that motion-induced corrections to the internal dynamics of a vertically moving atom exceed the ones obtained for the parallel-motion scenario by an order of magnitude \cite{Klatt2016}, we set out to study whether such qualitative changes may likewise be found for the quantum friction force. After all, the physics leading to both phenomena is identical. Looking at the results summarized in Table~\ref{tab:results}, we can conclude that there are indeed qualitative changes in the velocity dependence of the force when changing from parallel to vertical relative motion of the atom. While up to second order in the atom-field coupling, the quantum friction force is exponentially small for parallel motion, it is found to scale quadratic with relative velocity in the vertical case. Moreover it is consistently found to do so in both the Markovian and the perturbative approach. If considering the terms of fourth order in coupling as well, the Markovian approach only predicts a change in the numerical prefactor of the force when changing from parallel to vertical motion, whereas time-dependent perturbation theory suggests indeed an increase by one order of magnitude. However, we must recognize that even for extremely confident parameter estimates such as $v\!=\!10^3\,\T{m}/\T{s}$, $\w_{10}\!=\!10\,\T{THz}$, $\Ga_1\!=\!10^9\,\T{s}^{-1}$ and $z_\TA\!=1\,\T{nm}$, the best-case-scenario, that is in case of the largest prediction for the quantum friction force among the ones given in Tab.~\ref{tab:results},
\begin{align}
F_\T{fric}\propto\left(\!\tfrac{v}{\w_{10}z_\TA}\!\right)^{\!2}F_\T{CP}\,,
\end{align}
the friction force only amounts to about 1\% of the static Casimir-Polder force and thereby continues to elude detection.

\end{section}

\begin{section}{Acknowledgments}

This work was supported by the DFG (Grants BU 1803/3-1476, GRK 2079/1), the Freiburg Institute for Advanced Studies and the LANL LDRD program. M.B.F would like to thank ANPCyT, CONICET, UBA, LANL and CNLS. We are grateful to G. Barton, C. Henkel,  F. Intravaia, F. Mazzitelli, V. Mkrtchian and S. Scheel for discussions.
\end{section}

\end{document}